# Parallel and Distributed Simulation: Five W's (and One H)

Gabriele D'Angelo

*Abstract*—A well known golden rule of journalism (and many other fields too) is that if you want to know the full story about something you have to answer all the five W's (Who, What, When, Where, Why) and the H (How). This extended abstract is about what is missing in parallel and distributed simulation and how this affects its popularity.

## I. Introduction

IN the last decades, the systems to be analyzed have become more and more complex. For example, in computer networks some very sophisticated protocols have been proposed and many networks are made by a huge number of nodes. The performance evaluation of such kind of systems is often based on simulation. Given these requirements and the large diffusion of parallel/distributed systems, we would expect Parallel And Distributed Simulation (PADS) to gain massive popularity: this is not the case.

**WHAT**: are the PADS techniques ready for prime time after all the research work that has been done? Some fields are strongly based on them (e.g. digital virtual environments, wargames and military simulation) while others see a limited diffusion (e.g. performance evaluation, online gaming). In other words, many are unwilling to dismiss the "old" (sequential) tools and to switch to modern ones, despite the very strong demand for scalability and speed. What is missing?

**WHEN**: two of the main goals of the last decades research work on PADS were: i) *make it fast*; ii) *make it easy to use*. Today, we can say that PADS, in some conditions, can be very fast. Above all, the research work on synchronization algorithms and data distribution management has allowed to increase the speed-up of simulation runs a lot. This is true when the simulation model is properly partitioned among the execution architecture, the appropriate synchronization algorithm is used (each one of them has its characteristics and limitations), and the execution architecture is fast, reliable and in most cases homogeneous (in terms of performance of each node). In other words, the execution speed of PADS is limited by its slowest component and therefore a strict control on the whole simulator and its execution architecture is absolutely necessary. In terms of usability, there is not much more to be added: PADS does not work "straight out of the box". The level of knowledge modelers are required is still too high. Some aspects such as causality constraints are hard to manage and understand.

**WHO**: let's now investigate the simulation users more in detail. It is quite obvious that PADS techniques are, in some cases, necessary (and provide a benefit) while in others they are not necessary at all (e.g. when PADS is slower than sequential). Therefore, each time the main question should be: what is the better choice? There are many possibilities: as sequential/parallel/distributed only refer to the high level approaches. Yet we must remember that each of them can be implemented using different execution architectures (e.g. multi-core CPUs, clusters, public of private clouds). Up to now, the whole problem is left to the simulation model developer (or the simulator user), that often is not a PADS expert. It feels like PADS tools are for initiates: a better approach would be to hide from the users all such technical details that should be on duty of software tools.

**WHERE**: it has been said for years that we have a parallel simulation when the execution nodes are connected by a low latency network (e.g. a bus), and conversely a distributed one when the latency is higher (e.g. a LAN, WAN or even Internet). This categorization is simple and clear but inadequate: in the real world the execution architectures are much more complex and heterogeneous in terms of hardware, software and runtime conditions. Nowadays, the typical architecture is made by some multi-core (and often multi-processor) hosts interconnected with some kind of network: it is quite heterogeneous and is going to be even more. The *"everything as a service"* approach that is at the basis of Cloud Computing is going to hit PADS. If in case of Private Clouds is possible to expect some sort of control on the execution environment, this is not is conceivable in Public Clouds (such as Microsoft Azure or Amazon EC2). In the latter, the user can only rely on some general Service Level Agreements (SLAs). In this even more complex situation, the simulator user is left alone: the most simulators and their underlaying technologies are unable to tackle it.

**WHY** is it so difficult to decide what is the best approach? Even the "sequential or PADS" choice is hard to make because it depends on a very large set of dynamic parameters that can be found in all the logical layers of the architecture, starting from the simulation model behavior and down to the hardware performances. All those parameters need a case-by-case evaluation. Furthermore, they can also change within any single simulation run, due to many different factors such as the semantic of the simulated model and the unexpected presence

G. D'Angelo is with the Department of Computer Science, University of Bologna, Italy. e-mail: g.dangelo@unibo.it

of background load in the execution architecture.

## II. HOW: ON THE ADAPTIVE APPROACHES

Let's start with a warning: the "silver bullet" does not exist, even in simulation. The last attempt in PADS to obtain a "one fits all" solution has produced the IEEE 1516 - High Level Architecture (HLA) standard, that is quite complex to use, lacks some basic features and has lead to many performance issues.

In our vision [1], the most of what's said above can be reduced, in simplest terms, to a *partitioning problem*. That is about decomposing the simulation model into a number of components and to properly allocating them among the execution units. This allocation procedure has at least two main goals to pursue: the computation load in execution architecture has to be kept approximately balanced while the communication overhead has to be minimized [2]. If both these requirements are satisfied then the execution is likely to be efficient. The hard part is that all of this has to be: i) *transparent to users*, ii) *dynamic and adaptive* (given that both the model behavior and the execution architecture conditions are not predictable). In other words, the runtime conditions are largely unpredictable and moreover the environment is dynamic and very heterogeneous. The direct consequence is that, in this case, all static (and analytical) approaches are not adequate.

What we propose is the partitioning of the simulated model in very small parts (referred to as entities). Each entity represents a tiny piece of the simulated model and interacts with other entities to implement the model behavior. In this way, the execution architecture that is composed of multiple nodes is nothing more than a set of containers for the **Simulated Entities** (**SE**). Each container is usually called a **Logical Process**, or **LP** for short. Under the usability viewpoint, this is a Multi Agent System (MAS) [3]. A paradigm that has been demonstrated very easy to use, solid and promising. About our proposal, it is worth noting that the SEs are not statically allocated on a specific LP, conversely SEs can be migrated with the aim to satisfy the partitioning requirements that have been previously described and to improve the runtime efficiency of the simulator.

More in detail, two main aspects have to be considered. Firstly, with respect to a sequential (i.e. monolithic) simulation, every PADS has to deal with a significantly higher communication cost (e.g. network latency and bandwidth limitations). Reducing this cost to the bare minimum is of main importance. Secondly, the simulator speed is bounded by its slowest component and therefore smart load-balancing strategies have to be implemented. To reduce the communication cost, the main strategy is to cluster the highly interacting SEs within the same LP. This clustering has the effect of increasing the use of low latency and high bandwidth networks (e.g. the RAM within the host) and conversely reducing the usage of very costly communication technologies (e.g. LAN, WAN, Internet). It is pretty obvious that the other side of the problem is that clustering all the SEs in the same LP is (usually) not a good load-balancing strategy. Moreover, if a LP is overloaded, then it is going to slow-down the whole simulator and therefore the clustering is of secondary importance with respect to load-balancing. In other words, the partitioning is a very dynamic optimization problem with multiple goal functions and with a lot of parameters with unpredictable values.

As said before, the use of analytical methods to tackle such problem is unrealistic in this case: we have to rely on heuristic methods. A set of heuristics is used to evaluate the simulator and the execution architecture step-by-step, and to decide if reallocations (of SEs) are necessary. It is worth nothing that migrations are not free. They have a cost that is mainly given by the "serialization" of state variables (in SEs) and their network transfer. If it is favorable to cluster some SEs in the same LP, then some reallocations will be managed. That will also happen in case of imbalances in the execution architecture. All of this has to be done for the whole simulation length, given that both the execution architecture runtime conditions and the simulated model behavior will change with time. Some special cases should be analyzed more in detail. For example, if the amount of computation required by the simulation model is so low that no parallelization is necessary, then the execution architecture should automatically shrink up to a single LP (that is, a sequential simulation). Conversely, in case of a large amount of work the communication cost will be balanced by the benefit of parallel execution and therefore the number of LPs in the simulation needs to be increased.

## III. ARTÌS AND GAIN+

In the last years, we pursued the approach described in the previous Section with the implementation of a new simulation middleware (called Advanced RTI System, ARTÌS) and the companion GAIA+ framework (Generic Adaptive Interaction Architecture) [4].

While this effort is still not at the end, a lot of different systems and scenarios have been evaluated to validate our work. For example, both wired and wireless communication environments have been investigated. Using some of the features previously described it has been possible to manage the fine grained simulation of complex communication protocols (e.g. IEEE 802.11) when in presence of a huge number of nodes (up to 1 million) [1]. In the wired case, we worked on the design and evaluation of gossip protocols in unstructured networks (e.g. scale-free, small-world, random) [5]. In all of these cases, we obtained a very good performance increase with respect to traditional simulation techniques.

## REFERENCES

[1] G. D'Angelo and M. Bracuto, "Distributed simulation of large-scale and detailed models." *International Journal of Simulation and Process Modelling (IJSPM)*, vol. 5, no. 2, pp. 120–131, 2009.
[2] X. Zeng, R. Bagrodia, and M. Gerla, "Glomosim: a library for parallel simulation of large-scale wireless networks," *SIGSIM Simul. Dig.*, vol. 28, no. 1, pp. 154–161, 1998.
[3] M. Wooldridge, *An Introduction to MultiAgent Systems*, 2nd ed. Wiley Publishing, 2009.
[4] "Parallel And Distributed Simulation (PADS) Research Group," http://pads.cs.unibo.it, 2011.
[5] G. D'Angelo and S. Ferretti, "Simulation of scale-free networks," in *Simutools '09: Proceedings of the 2nd International Conference on Simulation Tools and Techniques*. ICST, Brussels, Belgium, Belgium: ICST (Institute for Computer Sciences, Social-Informatics and Telecommunications Engineering), 2009, pp. 1–10.